\documentclass[english,aps,prl,superscriptaddress,
               twocolumn,showpacs,amsmath,amssymb]{revtex4-1}
\usepackage{amsmath}
\usepackage{bm}
\usepackage{array}
\usepackage{graphicx}
\usepackage[margin=1.5cm]{geometry}
\usepackage{color}
\usepackage{amsfonts}
\usepackage{amssymb}

\begin{document}

\title{Frustrated bearings}

\author{R.~S.~Pires}
\affiliation{Departamento de F\'{\i}sica, Universidade Federal do Cear\'a,
    60451-970 Fortaleza, Cear\'a, Brazil}
\author{A.~A.~Moreira}
\affiliation{Departamento de F\'{\i}sica, Universidade Federal do Cear\'a,
    60451-970 Fortaleza, Cear\'a, Brazil}
\author{H.~J.~Herrmann}
\affiliation{Departamento de F\'{\i}sica, Universidade Federal do Cear\'a,
    60451-970 Fortaleza, Cear\'a, Brazil}
\affiliation{PMMH, ESPCI, CNRS UMR 7636, 7 quai St Bernard, 75005 Paris, France}
\author{J.~S.~Andrade~Jr.}
\email{soares@fisica.ufc.br}
\affiliation{Departamento de F\'{\i}sica, Universidade Federal do Cear\'a,
    60451-970 Fortaleza, Cear\'a, Brazil}

\date{\today}

\begin{abstract}
  In a bearing state, touching spheres (disks in two dimensions) roll on each
  other without slip. Here we frustrate a system of touching spheres by imposing
  two different bearing states on opposite sides and search for the
  configurations of lowest energy dissipation. If the dissipation between
  contacts of spheres is viscous (with random damping constants), the angular
  momentum continuously changes from one bearing state to the other. For Coulomb
  friction (with random friction coefficients) in two dimensions, a sharp line
  separates the two bearing states and we show that this line corresponds to
  the minimum cut. Astonishingly however, in three dimensions, intermediate
  bearing domains, that are not synchronized with either side, are energetically
  more favorable than the minimum-cut surface. Instead of a sharp cut, the
  steady state displays a fragmented structure. This novel type of state of
  minimum dissipation is characterized by a spanning network of slipless
  contacts that reaches every sphere. Such a situation becomes possible because
  in three dimensions bearing states have four degrees of freedom.
\end{abstract}

\maketitle
  A bearing is a set of spheres (discs in 2d) that, with the position of their
  centers fixed, roll on each other without any slip at their
  contacts~\cite{Herrmann1990, Manna1991, Baram2004}, transferring torque
  without dissipation. A sufficient condition to achieve a bearing state is to
  ensure that the graph of contacting spheres is bipartite, that is, all loops
  in the graph have an even number of spheres~\cite{Baram2004,Stager2016}. In
  two dimensions, the tangential velocities $v$ at all contacts must be
  identical, so that the bearing state is uniquely defined by $v$. In three
  dimensions, other types of bearing states can be identified~\cite{Stager2016}.
  
  The concept of bearings plays an important role on the dynamics of dense
  packings of particles~\cite{Alonso-Marroquin2006, Ayer1965, Jodrey1985,
  Yu1988, Bessis1990, Konakawa1990, Standish1991, Roux1993, Yu1993,
  Anishchik1995, Elliott2002, Baram2004, Ciamarra2006, Astrom2008, Halsey2009,
  Sobolev2010, Reis2012, Martin2015, Stager2016, Peshkov2019}. Bearings obtained
  by construction that completely fill space~\cite{Manna1991a, Oron2000,
  Baram2004a, Baram2005, Kranz2015, Stager2018} can support large pressures
  while allowing for sliding movement. Moreover, it was shown that the
  synchronization process necessary to reach a global bearing state can be
  substantially enhanced by adjusting the inertial contribution of individual
  rotors~\cite{Araujo2013}. When the particles move, inducing a complex and
  changeable force network~\cite{Tkachenko1999}, as it is the case in shear
  bands, bearing states form spontaneously~\cite{Astrom2000,Latzel2003}. Due to
  these properties, it was suggested~\cite{Herrmann1990, Roux1993,
  Astrom2000} that such bearing states may explain the existence of ``seismic
  gaps'', namely, regions in tectonic faults that should be moving, but where no
  earthquake activity has been detected for a long time~\cite{Lomnitz1982,
  Kagan1991}.

  While the transfer of momentum through disordered systems has been studied
  extensively, much less is known about the transfer of torque. In particular,
  when contacts dissipate due to Coulomb friction instead of viscous forces,
  systems can get stuck in particular configurations. Here we investigate a
  system of touching rotors subjected to frustrating boundary conditions,
  that is, systems where one side is forced to be in one bearing state, while on
  the other side another bearing state is imposed. Between these two bearing
  states there must be slipping contacts with random friction coefficients where
  energy is dissipated. Following its natural dynamics, a bearing system will
  settle to one configuration, albeit frustrated, of minimum energy loss.

  In our computational model, we place touching spheres (disks in 2d) on a
  regular grid in such a way that their positions are fixed but they can rotate.
  At every contact a tangential friction force $F_{t(ij)}$ acts on the pair of
  particles $i$ and $j$. We consider two cases, namely, either this force is
  viscous,
  \begin{equation}
    F_{t(ij)} = \eta_{ij} v_{r(ij)},
  \end{equation}
  where $\eta$ is a damping coefficient, and $\vec{v}_{r}$ is the relative
  slipping velocity
  \begin{equation}
    \vec{v}_{r(ij)} = \vec{\omega}_j\times \vec{r}_{ji}
                      - \vec{\omega}_i\times \vec{r}_{ij},
    \label{eqqq}
  \end{equation}
  or Coulomb-like
  \begin{equation}
    F_{t(ij)}=
   \left\{\begin{matrix}
    F_{s(ij)} & \text{if }v_{r(ij)}\ne 0 \\
   \mu_{ij} F_n & \text{if }v_{r(ij)}= 0,
   \end{matrix} \right.    
  \end{equation}
  where $F_s$ and $F_n$ are the static friction and the normal force on the
  contact, respectively. In order to find the stationary state, we use two
  different numerical techniques. On one hand, we find the configuration that
  minimizes the dissipation power $P$, which is defined as,
  \begin{equation}
    P=\frac{1}{2}\sum_i^{N}\sum_j^{\{ i\}}
      \vec{F}_{t(ij)}\cdot \vec{v}_{r(ij)},\label{Eq::P}
  \end{equation}
  where the sum in $j$ goes over the set $\{i\}$ of disks that are in contact
  with disk $i$. We use about $10^6$ iterations steps of gradient
  descent~\cite{Ruder2016} in order to find the state that minimizes
  $P=P(\vec{\omega}_1,\vec{\omega}_2,\ldots,\vec{\omega}_n)$. For a viscous
  friction force, the components of
  $\nabla_{\vec{\omega}_1,\vec{\omega}_2,\ldots,\vec{\omega}_n} P$ can be
  written as,
  \begin{equation}
    \frac{\partial P}{\partial \omega_{i,k}}=2\sum_{j}\eta_{ij}
    (\vec{\omega}_i\times \vec{r}_{ij}-\vec{\omega}_j\times
    \vec{r}_{ji})\cdot (\hat{k}\times \vec{r}_{ij}),
  \end{equation}
  while for a Coulomb-like friction force they become,
  \begin{equation}
    \frac{\partial P}{\partial \omega_{i,k}}=F_n\sum_{j}\mu_{ij}\frac{
      (\vec{\omega}_i\times \vec{r}_{ij}-\vec{\omega}_j\times \vec{r}_{ji}) }{
      \left \|\vec{\omega}_i\times \vec{r}_{ij}
      - \vec{\omega}_j\times \vec{r}_{ji}\right \| }
      \cdot (\hat{k}\times \vec{r}_{ij}),
  \end{equation}
  where the sum in $j$ goes just over the slipping contacts of sphere $i$. In a
  second approach, we use a Cundall-Strack scheme~\cite{Cundall1979,Brendel1998}
  in order to obtain an approximation for the friction force,
  \begin{equation}
   \vec{F}_t=\left\{\begin{matrix}
   \mu_k F_n(\vec{v}_r/||\vec{v}_r||) &
                \hspace{2cm} & \text{if }
                ||\vec{F}_s^*|| > \mu_s ||\vec{F}_n|| \\
   \vec{F}_s^* & \hspace{2cm} & \text{otherwise}.
   \end{matrix}\right.
  \end{equation}
  Here, $\mu_s$ and $\mu_k$ are the static and the dynamic friction
  coefficients, respectively, and $\vec{F}_s^*$ mimics the static friction
  force,
  \begin{equation}
   \vec{F}_s^*=-K_t\vec{\delta}-A_t\vec{v}_r.
  \end{equation}
  \begin{figure}[!t]
    \includegraphics[width=0.8\columnwidth]{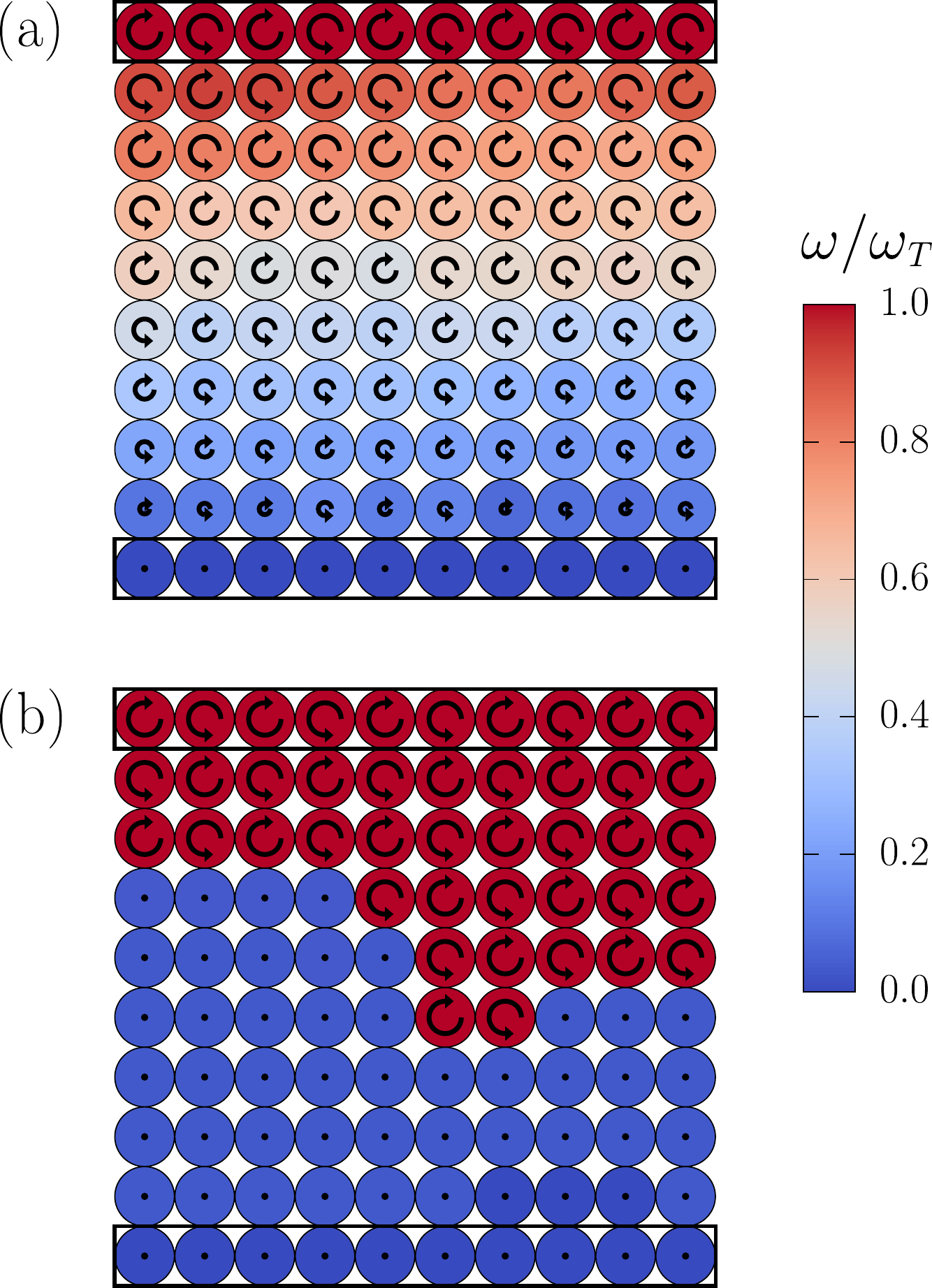}
    \caption{Stationary configurations for (a) viscous damping and (b) Coulomb
    friction in the contacts of a packing of disks whose centers are the
    vertices of a square lattice of size $10\times 10$. The boxes indicate the
    positions of the imposed boundary conditions. Disks at the bottom are
    constrained to remain static, $\omega_{B}=0$, while those at the top rotate
    with a given angular velocity, $\omega_{T}>0$, as a fixed bearing. Except
    for the bottom and top rows, all angular velocities of the disks were
    initially set randomly. Regardless of the initial condition, given the
    disorder in the friction coefficients, the system always evolves to the same
    stationary configuration. The color bar corresponds to the ratio between the
    moduli of the angular velocities $\omega$ of the spheres and the imposed
    modulus of the angular velocities at the top boundary $\omega_{T}$. There is
    no slipping or dissipation in the contacts between disks of the same color.
    In (b), the interface between the two bearing states is identical to the
    minimum cut corresponding to the friction coefficients of the contacts. This
    equivalence is due to the fact that dissipation occurs only at the
    interface. Simulations with 1000 different sets of disordered friction
    coefficients showed that the boundary corresponds to the minimum cut in 983
    cases. The remaining cases corresponded to systems with two distinct minimal
    cuts with values very close to each other. In such cases, the final state
    can present two boundaries at these cuts.
    \label{stat_conf4}
  }
  \end{figure}
  Static friction is idealized as an imaginary tangential spring used to keep
  the contact point slipless, where $\vec{\delta}$ is the elongation of this
  spring, $K_t$ its stiffness, and the constant $A_t$ damps any oscillation in
  the synchronized contacts. For simplicity, we use a viscous damping force
  proportional to the relative tangential velocity at the contact point
  $\vec{v}_r$. In our simulations $K=10^5$, $A=10^3$ and $F_n=50$ for discs
  (spheres in 3D) of radii 0.5. These values ensure that for very small
  $\delta=10^{-4}$ the threshold of dynamic friction is reached. We tested
  different values of $F_n$ and greater values of $K$ and $A$, but no
  significant differences were observed.
  
  For the elongation $\vec{\delta}$, we use the following representation, 
  \begin{equation}
   \vec{\delta}=\left\{\begin{matrix}
   -(\mu_k ||\vec{F}_n||/K_t)\vec{F}_s^* &
         \text{if } || \vec{F}_s^* || > F^*_{s\;\text{max}}=\mu_s ||\vec{F}_n|| \\
   \displaystyle\vec{\delta}(t^\prime) + 
         \int_{t^\prime}^{t}{\vec{v}_r}{dt^{\prime\prime}} & \text{otherwise},
   \end{matrix}\right.
  \end{equation}
  where $t^\prime$ is the time when $\vec{v}_r$ changes for the first time after
  $F_s^*$ reaches the threshold
  $F^*_{s\;\text{max}}$~\cite{Cundall1979,Brendel1998}. We use Gear's
  algorithm~\cite{Gear1967, Allen1987} to integrate the equations of motion,
  \begin{equation}
    I\frac{\partial^2 \vec{\omega}_i}{\partial t^2}=\sum_{j}^{\{i\}} \vec{r}_{ij} \times \vec{F}_{t(ij)}, 
  \end{equation}
  together with the differential equation for the elongation.

  The Cundall-Strack approach provides a way to solve the equations of motion
  and simulate the evolution of the system. The gradient descent finds the
  state of minimal dissipation without giving information on the dynamics. The
  advantages are that it allows simulations of larger systems and discriminate
  slipless contacts. In all cases, the system always reached the same final
  state regardless of the initial conditions or the method employed, suggesting
  that the dynamics naturally selects a unique state of minimum
  dissipation~\cite{Stegmann2011}. This observation is compatible with the
  minimum entropy production principle~\cite{Prigogini1978}.
    
  In all simulations with viscous friction, the damping coefficients between
  contacts are randomly chosen according to a uniform distribution in the
  interval $0.1\leq \eta \leq 1.0$. For simulations with Coulomb-like friction,
  the coefficients $\mu=\mu_{s}=\mu_{k}$ are also randomly chosen uniformly
  between $0.1$ and $1.0$ for each contact between spheres (disks in 2d). We
  maintain the magnitude of the angular velocities of the top $\omega_{T}$ and
  bottom $\omega_{B}$ planes (rows in two dimensions) of the system in different
  fixed bearing states. In two dimensions, the disks at these rows in even and
  odd columns spin in opposite directions, assuring that their contacts are
  slipless. In three dimensions, the bearing states are achieved by imposing
  that the spheres at the top and bottom planes spin with angular velocities
  that point to a $45^\circ$ diagonal direction $\pm(\hat{x}+\hat{y})$
  orthogonal to the vertical direction $\hat{z}$.
  
  We first consider a two-dimensional bearing where the centers of the discs are
  put on a square lattice, as shown in Fig.~\ref{stat_conf4}. The disks of the
  top row are kept rotating, $\omega_{T}>0$, while those at the bottom are
  static, $\omega_{B}=0$.
  \begin{figure}[!t]
    \includegraphics[width=\columnwidth]{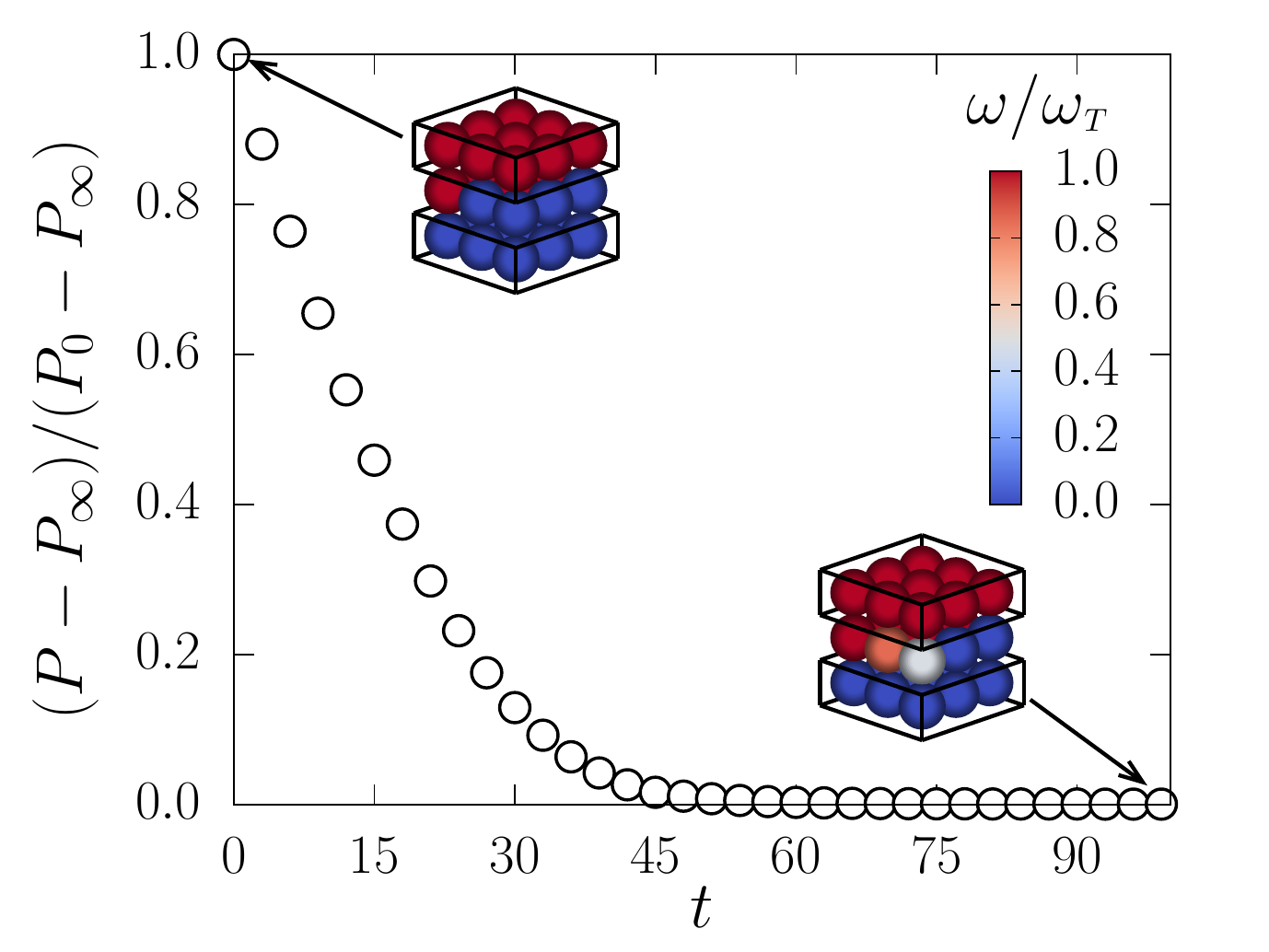}
    \caption{Dissipation power as a function of time for a cubic packing of
    touching spheres of size $3\times 3\times 3$ with Coulomb friction in the
    contacts, modeled by Cundall-Strack. The spheres at the bottom plane are
    constrained to remain static, $\omega_{B}=0$, while those at the top plane
    rotate with a given angular velocity, $\omega_{T}>0$, as a fixed bearing
    state. As in the 2d case, here the system evolves to the state of minimum
    dissipation. In 3d, however, the energy dissipation becomes smaller than
    when one just splits the system in two domains by the minimum-cut surface,
    and in the stationary state some spheres do not belong to either of the two
    fixed bearings. Here $P_{0}$ and $P_{\infty}$ are the initial and stationary
    dissipation powers, respectively. We also show snapshots of the system, one
    at the beginning and one at the end of the evolution. The boxes indicate the
    positions of the imposed boundary conditions.
    \label{brg3x3x3}
    \label{Pt3x3x3}}
  \end{figure}
  After a long time, the system reaches a stationary state which, for the Coulomb
  case, corresponds to a global minimum in dissipated energy since it always
  decreases monotonically in time. In Fig.~\ref{stat_conf4}, we compare viscous
  damping and Coulomb friction. While in the first case the angular velocities
  change continuously from top to bottom, for Coulomb friction the system splits
  in two different fixed bearing states, where the initially free discs in the
  middle finally follow either the rotating bearing state on the top or the
  fixed bearing state on the bottom. Between the red and blue regions of
  Fig.~\ref{stat_conf4} emerges a line of slipping contacts separating the two
  fixed bearing states.

  In the observed stationary states, we find only two domains, one synchronized
  with each boundary condition. Therefore, in the interface, all slipping
  contacts have the same relative velocity $\vec{v}_{r(ij)}$. Consequently, the
  dissipation power $P$ is proportional to the sum of the dynamic friction
  coefficients in the slipping contacts. We can define a network of contacts
  where the vertices are the centers of the disks, and an edge exists between
  any contacting disks. In 2d there is a dual network with vertices in the gaps
  between spheres. Interestingly, the line of slipping contacts is
  \emph{exactly} the minimum cut or shortest weighted path in the dual network,
  considering as weights of the bonds the friction coefficients of the
  corresponding contacts, for any distribution of random weights. The shape of
  the interface should only depend on the friction coefficients and not on the
  type of bearing states imposed externally. This is indeed the case if one
  replaces the lower fixed spheres by another non-static bearing state. We can
  see a close connection of this problem with the max-flow min-cut
  theorem~\cite{Elias1956,Ford1956,Goldberg1988}.
  \begin{figure}[!t]
    \includegraphics[width=\columnwidth]{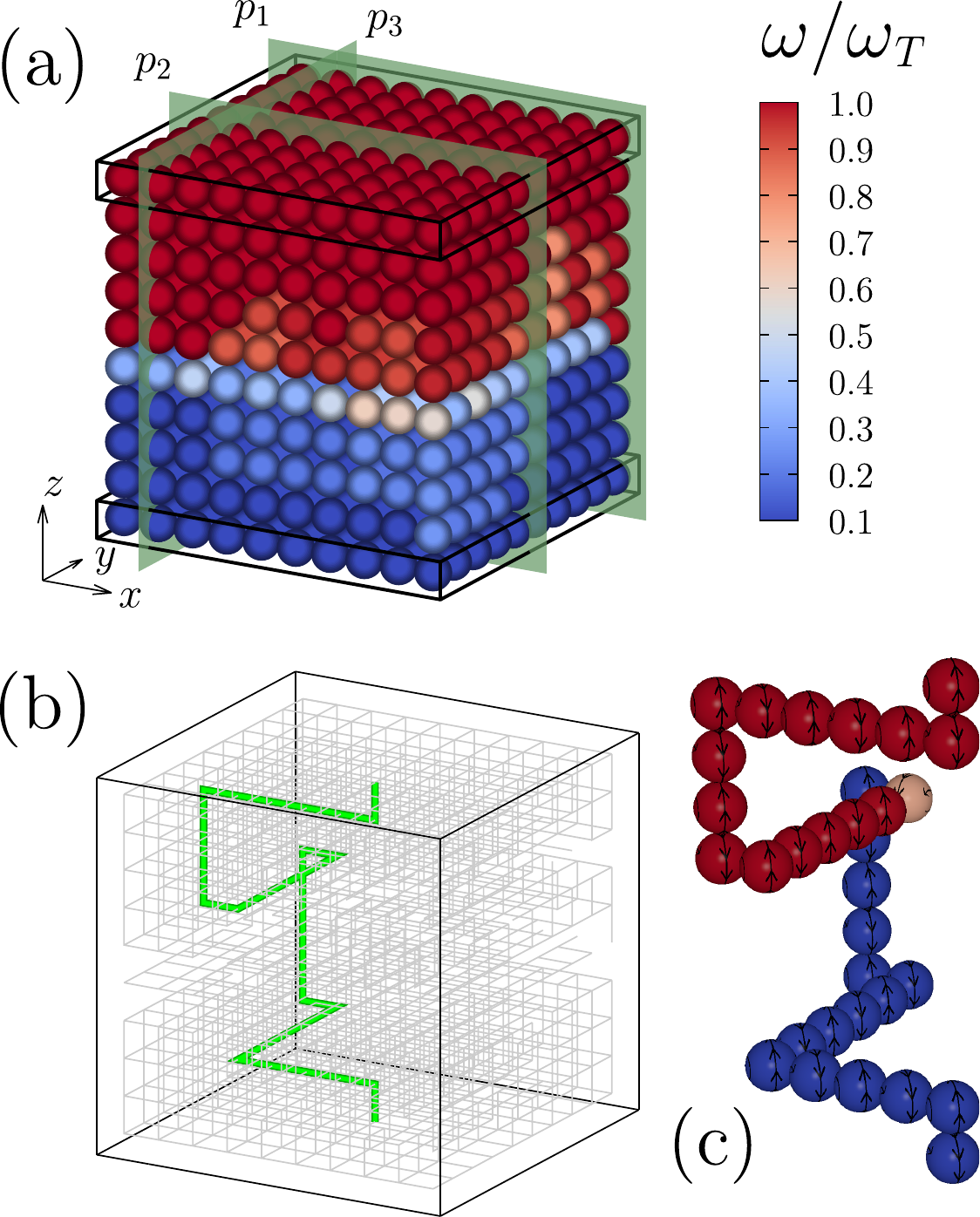}
    \caption{
    (a) Configuration of minimum dissipation of a cubic packing of size
    $10\times 10 \times 10$ with Coulomb friction in the contacts, obtained with
    the gradient-descent method. The boxes at the bottom and top indicate the
    imposed boundary conditions. The spheres at the top and bottom planes are
    spinning as fixed bearing states with $\omega_{T}=10\omega_{B}$. (b) Network
    of slipless contacts (in gray) of the configuration shown in (a). The
    highlighted green path, composed of slipless contacts, connects the top and
    bottom boundaries. In (c) the spheres along the same path are shown with
    their corresponding three-dimensional rotations.
    \label{brg10x10x10}}
  \end{figure}

  In three dimensions, we also observe for viscous damping in the stationary
  state a continuous transition between the two fixed bearing states. In
  Fig.~\ref{brg3x3x3}, we show the dissipation power as a function of time for a
  $3\times 3\times 3$ system of spheres with Coulomb friction in their contacts.
  As opposed to two dimensions, we do not find in the stationary state a single
  surface dividing two different fixed bearing states, but many spheres that do
  not belong to either of them. In fact, even starting with an initial condition
  in the minimum-cut configuration, the system evolves towards a stationary
  state that has a smaller dissipation power. In other words, in the 3d case,
  the minimum cut in the network of friction coefficients is generally not the
  state of lowest possible dissipation.
    
  The same behavior is observed for larger systems. In Fig.~\ref{brg10x10x10}a,
  we see a $10\times 10\times 10$ cubic bearing, where many spheres in the
  middle of the bearing are in intermediate bearing states. In this case, the
  spheres at the top and bottom planes are spinning as fixed bearing states with
  $\omega_{T}=10\omega_{B}$. For a $32 \times 32 \times 32$ bearing system, the
  distribution of power dissipation at the contacts ranges over eleven orders of
  magnitude (see Supplemental Material~\cite{SupMat}), although $0.1\le \mu \le
  1.0$.
  \begin{figure}[!t]
    \includegraphics[width=0.8\columnwidth]{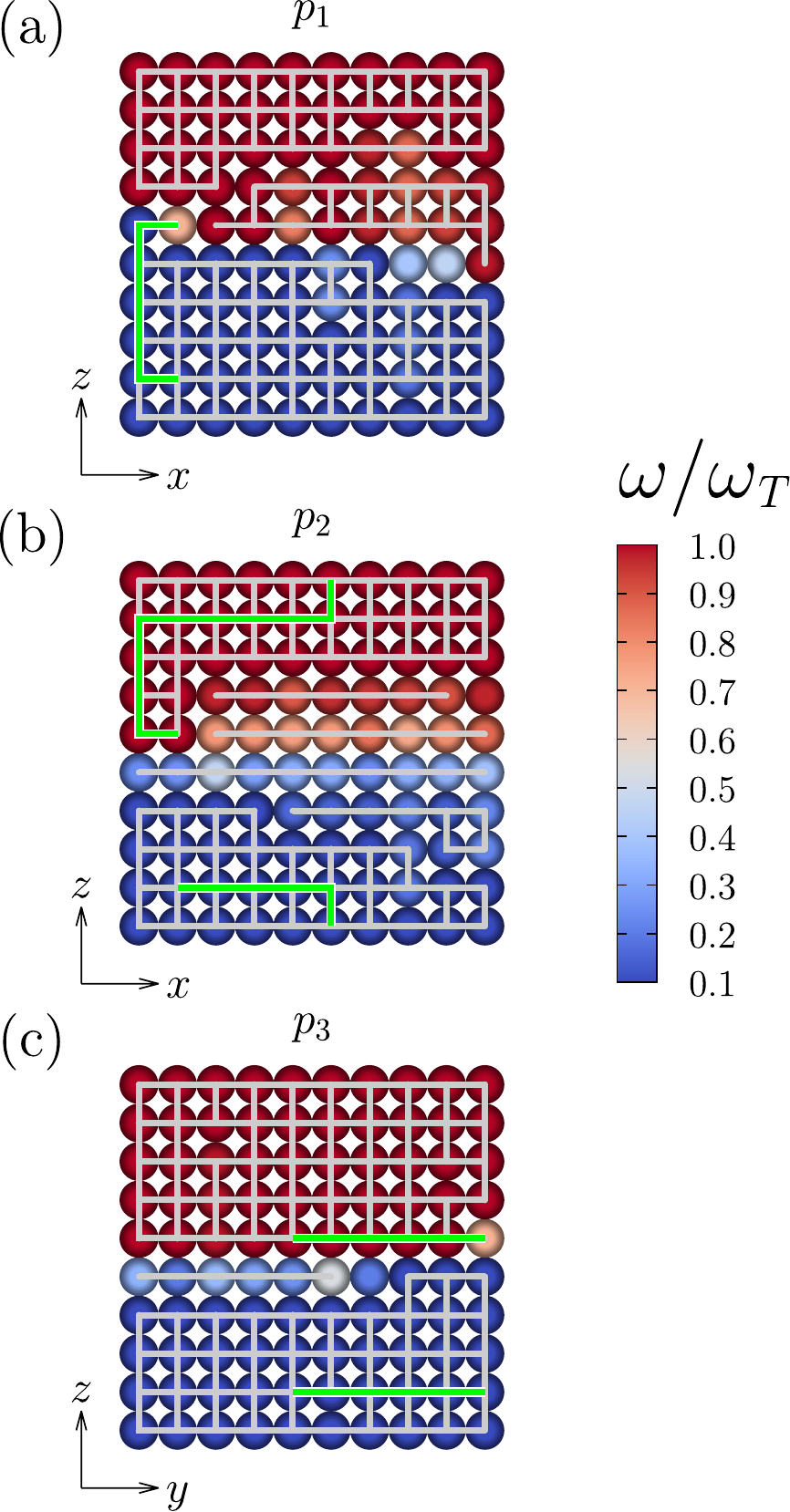}
    \caption{Cuts of the configuration along the three green planes in
    Fig.~\ref{brg10x10x10}a. The gray lines correspond to slipless contacts and
    the green lines correspond to the pieces composing the path, of slipless
    contacts, that connects bottom and top boundaries, as shown in
    Fig.~\ref{brg10x10x10}b. \label{cut10x10x10}}
  \end{figure}
  In this case, due to the large number of possible configurations, we could not
  determine the exact minimum-cut surface. However, as in the $3\times 3\times
  3$ case, Fig.~\ref{brg10x10x10}a shows that, in the state of minimum
  dissipated power, some spheres have intermediate angular velocities.
  
  At this point, two relevant questions arise. First, how do these states with
  dissipation lower than the minimum-cut configuration appear? Second, what are
  the distinctive properties of these newly discovered stationary states of
  minimum dissipation? We identified for the configuration shown in
  Fig.~\ref{brg10x10x10}a all slipless contacts and discovered that their
  network spans throughout the system, namely, the two opposite fixed bearing
  states are connected by paths of slipless contacts as shown in
  Fig.~\ref{brg10x10x10}b (gray lines). In Fig.~\ref{brg10x10x10}b we show one
  of these paths (highlighted in green), while the spheres along the same path
  are depicted in Fig.~3c with their corresponding three-dimensional rotations.
  In two dimensions, as shown in Fig.~\ref{stat_conf4}b, we have slip along the
  entire minimum cut, so that such connecting paths of slipless contacts cannot
  exist. Therefore a connecting path of slipless contacts, like the green one
  shown in Fig.~\ref{brg10x10x10}b, must go through several planes, as shown in
  Fig.~\ref{cut10x10x10}. The presence of these paths in the system somehow
  prevents the global frustration imposed by the fixed bearing states at the
  bottom and top boundaries. There are a few links through which most connecting
  paths go, that is, these links have large in-betweeness. In addition, we
  observe that the network of slipless contacts attains practically every
  sphere, as depicted in Fig.~\ref{brg10x10x10}b. Changing the imposed bearing
  states at the boundaries to other bearing states does not change the network
  of slipless contacts, showing that, as in two dimensions, it only depends on
  the disorder in friction coefficients.  
  
  The main reason for the difference between two and three dimensions is that in
  2d the bearing state has only one degree of freedom, namely, the tangential
  velocity, while in 3d there are four independent degrees of
  freedom~\cite{Baram2004}. In three dimensions, the tangential velocities
  $\vec{v}_{r(ij)}$ at slipping contacts depend on the sum of two vector
  products, Eq.~(\ref{eqqq}), which implies a coupling between the components of
  the angular velocities.

  In summary, we found that while for viscous damping there is a continuous
  change between bearing states, for Coulomb friction jumps appear: In two
  dimensions, a sharp interface separates the two bearing states, which is
  identical to the minimum cut. In three dimensions, we discovered a new type of
  final state in which the network of contacts without slip spans from one fixed
  bearing state to the other, attaining practically every sphere of the system.
  Our frustrated bearing in three dimensions with Coulomb friction is an example
  of a new kind of separation or fracture in a system consisting of an entire
  set of surfaces and fragments in between, reminiscent of shear failure of
  rocks under moderate confining pressure. As a future challenge, it would be
  interesting to study the bottlenecks and their in-betweeness in more detail.

  We thank the Brazilian agencies CNPq, CAPES, FUNCAP, and the National
  Institute of Science and Technology for Complex Systems (INCT-SC) in Brazil
  for financial support.

\end{document}